\documentclass[12pt,preprint]{aastex}

\newcommand{\axaf}{\mbox{\em Chandra\/}}
\usepackage{graphicx}
\usepackage{fancyheadings}
\usepackage{color}

\shorttitle{X-ray Jet J0841+1311 at z=1.866}
\shortauthors{D. A. Schwartz et al.}

\begin{document}

\title{Discovery of a Jet-Like Structure at the High Redshift QSO
  CXOMP J084128.3+131107}
\author{D.A. Schwartz, J. Silverman, M. Birkinshaw \altaffilmark{1},
  M. Karovska, 
  T. Aldcroft, W. Barkhouse, P. Green, D.-W. Kim,  B. J. Wilkes,
  D. M. Worrall\altaffilmark{1}}
\affil{Harvard-Smithsonian Center for Astrophysics, Cambridge, MA 02138}
\altaffiltext{1}{also, University of Bristol}

\email{das@head.cfa.harvard.edu}
%\slugcomment{DRAFT, \today, das}

\begin{abstract}

The Chandra Multiwavelength Project (ChaMP) has discovered a jet-like
structure associated with a newly recognized QSO at redshift
z=1.866. The system was 9.4\arcmin\ off-axis during an observation of
3C 207. Although significantly distorted by the mirror PSF, we use
both a raytrace and a nearby bright point source to show that the
X-ray image must arise from some combination of point and extended
sources, or else from a minimum of three distinct point sources.  We
favor the former situation, as three \emph{unrelated} sources would
have a small probability of occurring by chance in such a
close alignment. We show that interpretation as a jet emitting X-rays
via inverse Compton (IC) scattering on the cosmic microwave background
(CMB) is plausible. This would be a surprising and unique discovery of
a radio-quiet QSO with an X-ray jet, since we have obtained upper
limits of 100 $\mu$Jy on the QSO emission at 8.46 GHz, and limits
of 200 $\mu$Jy for emission from the putative jet.

\end{abstract}

\keywords{quasars: general --- galaxies: jets --- X-rays: galaxies}

\section{INTRODUCTION}

The objectives of the \axaf\ Multiwavelength Project (ChaMP) include
identification and categorization of a complete, well-defined sample of
serendipitous sources \citep{Kim03, Green03}. The results will be of
use, e.g., to study luminosity functions and their evolution, to
quantify the newly resolved source(s) of the hard diffuse X-ray
background, and to study cosmic structure and clustering of AGN and
galaxies. The wide angle nature of this survey also makes it ideal to
discover rare and unusual objects suitable for detailed study; e.g.,
lensed QSOs and X-ray jets.

Schwartz (2002a,b) has pointed out that if the jets observed in X-rays
on scales of tens to hundreds of kpc are emitting via IC scattering of
the CMB as suggested by \citet{Tavecchio00} and \citet{Celotti01},
then they will maintain the same apparent surface brightness
independent of redshift, and therefore can be detected to arbitrarily
large redshifts, up to the epoch at which they form.  The
\emph{Chandra} observations of such large scale jets in QSOs and
powerful FR II radio sources are typically interpreted as IC/CMB
emission, \citep{Schwartz00, Harris02,
Marshall01,Sambruna01,Siemiginowska02}.  All such interpretations
require the assumption that the jet is either relativistically beamed
with Doppler factors of order $\delta \sim$ 3 to 15, or that the
energy density in relativistic electrons grossly exceeds the magnetic
field energy density by at least two orders of magnitude. Detection of
the X-ray ``beacons'' predicted by Schwartz (2002a,b) would provide
additional evidence that the above assumptions are well founded.

We report the discovery of a candidate for such a system: CXOMP
J084128.3+131107, (hereafter called J0841). The X-ray image shows an
elongated structure. Despite the broad point response function (PSF)
of the \axaf\ telescope at this 9.4\arcmin\ off-axis angle, we show
that at least three point sources would be required to simulate the
observed extent. We favor an interpretation of emission 
from the jet of an optically identified QSO which is close to
the peak X-ray intensity.  We also mention alternate interpretations. Due to
the small probability for three \emph{unrelated} sources to occur by
chance in this configuration, such interpretations may be even more
unusual.

\section{OBSERVATIONS OF J0841}
\label{sec:obs}

\begin{figure}[h]
%\begin{minipage}[c]{0.46\textwidth}
%\includegraphics*[width=\textwidth]{f1.eps}
%{corejet2pixels50min.25to1.25BW.ps}
\plotone{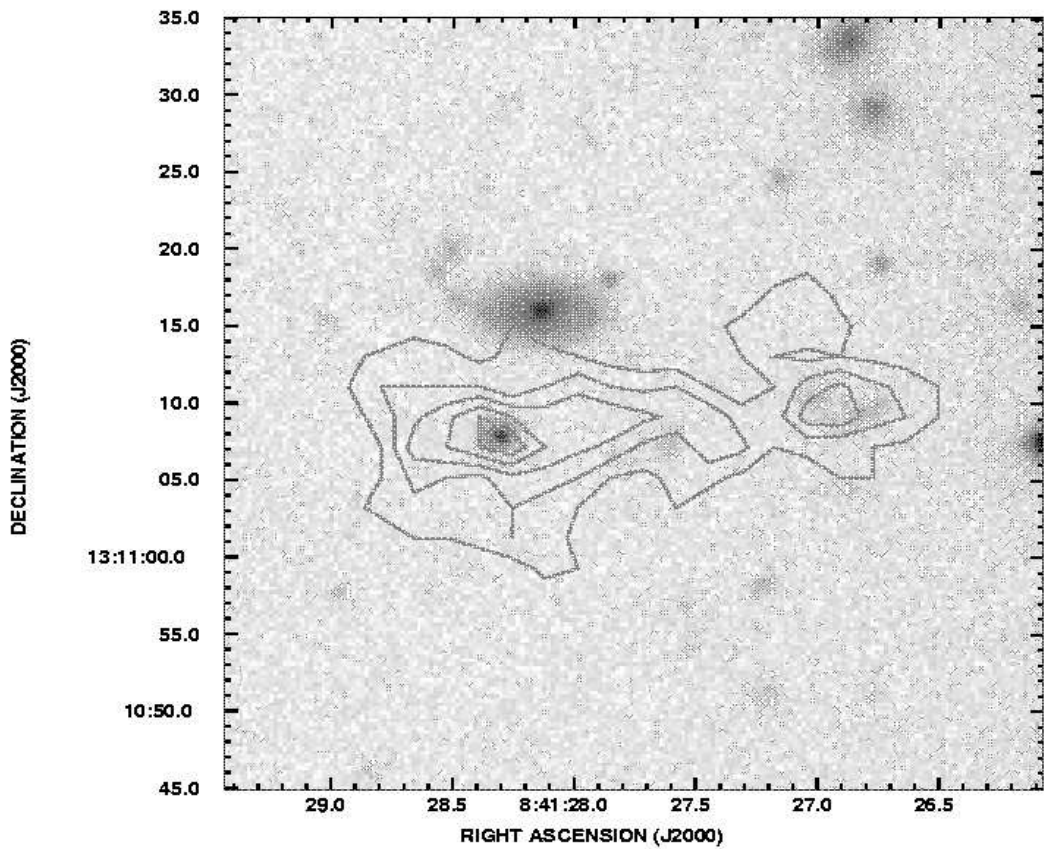}
\caption{\label{fig:cont} X-ray contours (0.5 to 7 keV) in the region
of J0841, superposed on a red-band image. Contour levels are 0.25,
0.50, 0.75, 1.0, and 1.25 counts per 0\farcs98$\times$0\farcs98
pixel. Background is 0.03 counts per pixel.  The r$^{\prime}$=20.9
object in the eastern contour peak is a QSO at redshift z=1.866.
The position difference between the X-ray peak and the optical source
is 1\farcs5, consistent with the \axaf\ PSF distortion at this large
off-axis angle.}
%\end{minipage}
\end{figure}

\begin{figure}[h]
\includegraphics*[width=3.in]{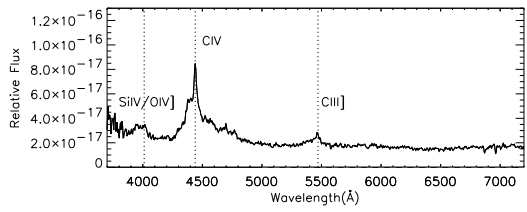}
%{j0841spectrumOpt.epsi}
\caption{\label{fig:spectrum} Ten minute Magellan exposure of J0841. The 
broad emission lines give a redshift 1.8661 $\pm$ 0.0005}
\end{figure}

The serendipitous detection of J0841 on the ACIS-I2 chip occurred
using the data from obsid 2130, an observation of 3C 207 with ACIS-S3
\citep{Brunetti02}.  Figure~\ref{fig:cont} shows the X-ray contours
superposed on a red-band image.  The strongest X-ray peak is
coincident within 1.\arcsec 5 with an r$^{\prime}$=20.9 mag object.  A
spectrum of this object (Fig.~\ref{fig:spectrum}) was obtained in a 10
minute exposure on Magellan using LDSS-2, and clearly shows a broad
emission line QSO. The optical data have about 13 \AA \,
resolution. The spectrum was cross-correlated against the composite 
SDSS QSO spectrum \citep{VandenBerk01} to give a redshift 1.866.

\clearpage

\clearpage

\begin{figure}[h] 
%\begin{minipage}[c]{0.46\textwidth}
%\includegraphics[viewport= 37 285  532  725,clip,width=\textwidth]{f3.eps}
%{j0841figps.ps
\plotone{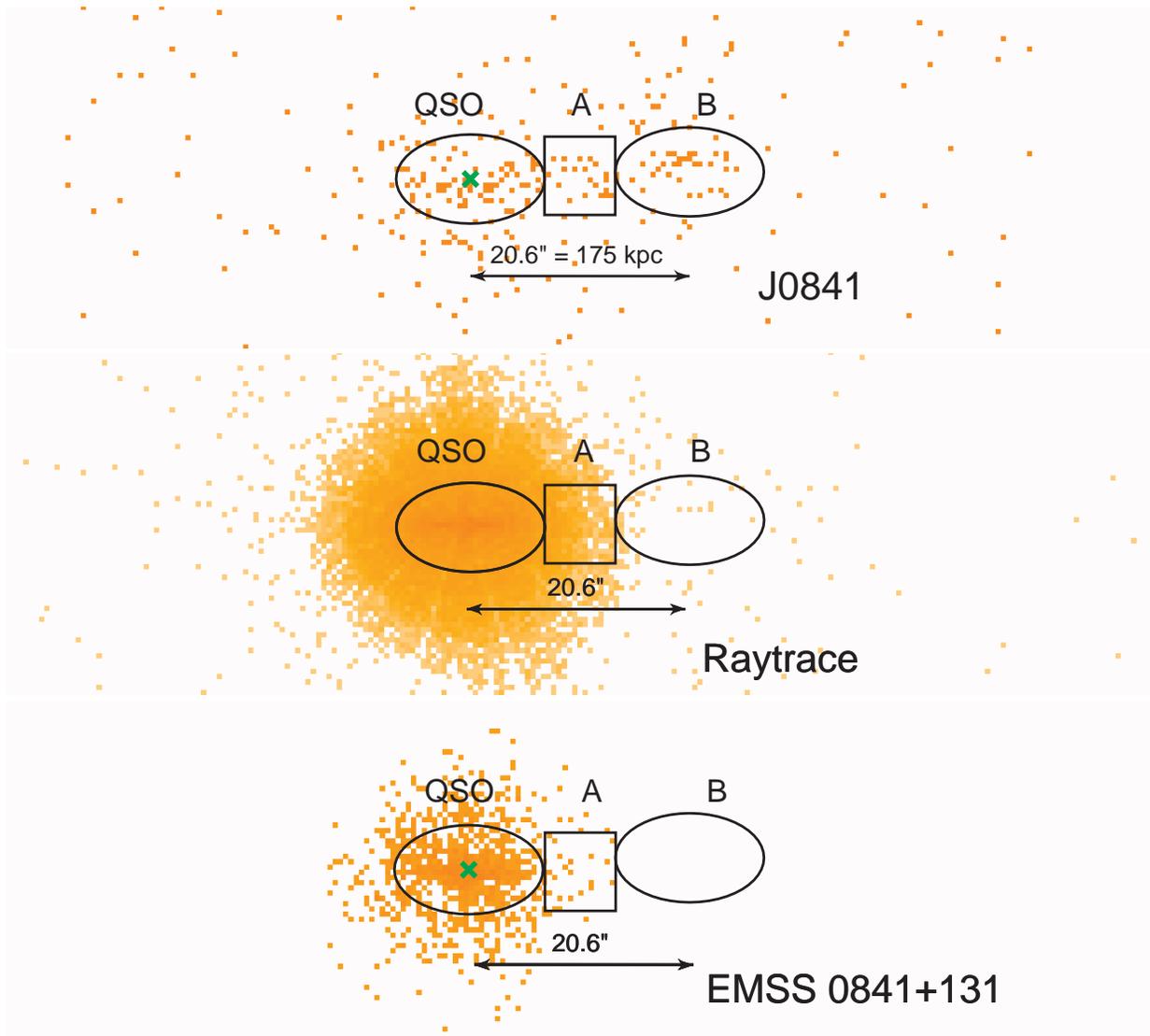}
%\vspace{-4cm}

\caption{\label{fig:images} X-ray images (0.5 to 7 keV) in 0\farcs49
  bins. Top to bottom, the J0841 system, a raytrace image of a 1.5 keV
  point source at the off-axis position of J0841, and a point source,
  EMSS 0841+1314, from the same observation as J0841. The ellipses
  labeled \emph{QSO} are centered on the QSO's (green crosses, top and
  bottom) and the raytrace axis (middle). The ellipse marked \emph{B}
  is placed on the centroid of counts associated with the
  concentration 10\arcsec above the right arrow in the top panel. It
  is then placed in the same relative position to the raytrace axis
  and EMSS 0841+1314, in the middle and bottom panels.  We show
  that box \emph{A} in the top panel has excess counts, and therefore
  represents a third source, based on ratios of the counts inside the
  \emph{QSO} ellipses to those inside the box \emph{A} in the lower
  two panels (see text).}

%\end{minipage}
\end{figure}

Although the contours in Fig.~\ref{fig:cont} seem to indicate an
extended X-ray structure, one must be careful due to the distorted
telescope response at this large off-axis angle.
Figure~\ref{fig:images} shows the X-ray data in the region of the QSO,
together with data around the nearby \emph{Einstein} medium survey
point source MS0838.6+1325 (\cite{Maccacaro91}, a z=0.723 QSO, also
called EMSS 0841+131), which happens to lie in the same \axaf\ field
at a similar off-axis angle, 9.\arcmin 3, and at the nearby azimuth of
247\arcdeg\ vs. 265\arcdeg\ for J0841. Each is compared with a high
fidelity raytrace\footnote{http://cxc.harvard.edu/chart} of a 1.5 keV
point source at this off-axis angle and the same azimuth as
J0841.\footnote{Note that the \emph{Chandra} point response function
is azimuthally asymmetric, \\ see http://cxc.harvard.edu/ccw.02} Both
QSOs are expected to have relatively hard spectra, for which 1.5 keV
is a good mean energy, so we do not expect significant effects due to
spectral differences. J0841 is clearly not a single point source.

We now show that two point sources could not produce the observed
X-ray structure.  Specifically, in the top panel of
Fig.~\ref{fig:images}, taking point sources at the QSO position and at
the center of the ellipse marked \emph{B}, we show that region
\emph{A} contains a significant excess of counts over background plus
those counts which could be attributed to the QSO, plus those counts
which could be attributed to the source \emph{B}. The expected counts
in box \emph{A} are based on the measured ratio of counts in
the ellipse marked \emph{QSO} to the counts in
a box marked \emph{A} to the west of the QSO, or a similar box to the
east of the QSO (not shown). We derive this predicted ratio both from real
data, EMSS 0841+1314, and from a raytrace, and in both cases we
predict $\le$10 counts in box \emph{A}, (including the non-X-ray
background).  However, we observe 21 counts in box \emph{A}, and the
probability of this is less than 0.1%.

We present the expected number of counts in box \emph{A} in more
detail for both methods: based on the raytrace image (middle panel of
Fig.~\ref{fig:images}), and based on the observation of EMSS 0841+131
(bottom panel of Fig.~\ref{fig:images}). The raytrace contains 1567
counts in the box \emph{A}, and 29945 in the QSO ellipse, for a
measured ratio of 0.052.  For EMSS 0841+131, after background
subtraction, those numbers are 28.3 and 669.6, for a ratio of
0.042$\pm$0.008, consistent with the raytrace prediction.  From the
observed 78.6 net counts inside the J0841 QSO ellipse (top panel of
Fig.~\ref{fig:images}), after background subtraction, we use the
raytrace result of 0.052 to predict 4.1 counts from the quasar would
fall in box \emph{A}.  We do a similar analysis, but with the raytrace
or EMSS 0841+131 source centered in the \emph{B} region.  We predict
0.113 and 0.091+/-0.012, respectively for the raytrace and for the
EMSS 0841+131 data, for the fraction of counts inside the \emph{B}
region which would appear in the box \emph{A}. From the net 32.6
counts observed inside region \emph{B}, (top panel of
Fig.~\ref{fig:images}), after background subtraction, the raytrace
predicts an additional 3.7 counts in region \emph{A} due to the point
source in region \emph{B}.  Thus for region \emph{A} in the top panel
of Figure~\ref{fig:images}, we measure 21 counts, and predict 7.8 from
the putative point sources \emph{QSO} and \emph{B}, plus 1.7
background counts.  The probability of observing 21 or more when 9.5
are expected is 0.086\%. We conclude that a minimum of 3 point sources
would be needed if J0841 does not have extended X-ray emission.

The ellipses drawn in Fig.~\ref{fig:images}) are 7\arcsec\ $\times$
4\farcs2, and are a contour of 62\% encircled energy based on EMSS
0841+131, or 55\% encircled energy based on the raytrace.  The
differences in these numbers are consistent with the statistics. For
this type of analysis we could have drawn any particular curve
around the QSO core -- the particular ellipse chosen was convenient
but arbitrary. The (unknown) true number of counts
is not relevant: we can predict that the
contributions to box \emph{A} from a true  total point source flux are
only about 2.9\% and 6.2\% from the west and east, respectively. 

There are about 100 sources deg$^{-2}$ above a flux of 10$^{-14}$ ergs
cm$^{-2}$ s$^{-1}$ \citep{Giacconi01}.  So there is a 2\% chance that
an unrelated source such as \emph{B} could occur within 30\arcsec\ of the
QSO.  There is then only about a 0.3\% chance of an independent
third source appearing in a 10\arcsec\ x 30\arcsec\ region between
the first two sources.  If we have three point sources, the
probability is $\le 6 \times 10^{-5}$ that they are
unrelated. However, the ChaMP survey will eventually find of order
10$^3$ QSOs brighter than r$^{\prime}$=21, so there might be as
large as 10\% probability for one such  system of unrelated
point sources to be found. 

\clearpage

\clearpage

\section{INTERPRETATION AS AN X-RAY JET}
In Figure~\ref{fig:images}, we will interpret the 78.6 net counts
measured in the
region indicated \emph{QSO} as from the QSO core, and the 32.6 counts
in region \emph{B} and the net 11.5 from region {A} as from the jet.
The  ellipses shown are 55\% encircled energy regions, based on the
raytrace result, giving an 
inferred total  counts of 143 from the QSO, and 80 from the jet.  This
total of 223  inferred  counts compares with 275 counts measured in a
25\arcsec\ radius circle about the QSO, which area contains an
expected 73.7 background counts.  The observation duration was 37542
seconds, (obsid 2130 of 3C 207). Taking a conversion of
6$\times$10$^{-12}$ ergs cm$^{-2}$s$^{-1}$ per count s$^{-1}$
(appropriate for an X-ray spectral energy index $\alpha$=0.7, and the
measured Galactic absorption n$_{\rm {H}}$=5$\times$10$^{20}$cm$^{-2}$
\citep{Stark92}) gives estimated measured fluxes of 2.3 x 10$^{-14}$
ergs cm$^{-2}$s$^{-1}$ for the QSO and 1.3 x 10$^{-14}$ ergs
cm$^{-2}$s$^{-1}$ for the jet, in the 0.5 to 7 keV band. At z=1.866
this gives luminosities\footnote{We use $H_0 = \rm 71\, km\, s^{-1}\,
Mpc^{-1}$ and a flat accelerating universe with $\Omega_0 = 0.27$, and
$\Omega_{\Lambda} = 0.73$.} of 5.7$\times$10$^{44}$ ergs s$^{-1}$ for
the QSO, and 3.2 $\times$10$^{44}$ ergs s$^{-1}$ for the jet. The
roughly 20\arcsec\ length of the jet on the sky corresponds to a minimum
length of 170 kpc at the redshift z=1.866.

Dividing the spectral data into six bins from 1 to 5 keV, and 
fixing the Galactic absorption, we can estimate an X-ray power-law
energy index of 0.3 $\pm$ 0.3 for the QSO and 0.5 $\pm$ 0.3 for the
jet region.

We made a 1 hour VLA observation in the C-array at 8.46 GHz on 10 Jan
2003, and find no emission from the QSO to a 3$\sigma$ rms noise limit
of 100 $\mu$Jy\footnote{See http://www.star.bris.ac.uk/\~{
}mb1/j0841.html}, or from the jet to a limit 200 $\mu$Jy.  The broad
band spectral indexes are $\alpha_{ox}$ = 1.43, and $\alpha_{ro} <
0.04,$ making it radio quiet, with a normal X-ray to optical ratio.
Although it would be extremely surprising, and unprecedented, for a
radio quiet QSO to have a jet, it can be reasonably interpreted if the
jet is highly beamed toward our line of sight, and if the X-rays are
being produced by inverse Compton (IC) scattering on the cosmic
microwave background (CMB).  This is due to the extra factor of
$\delta^{1+\alpha}$ \citep{Dermer94} by which the X-rays are boosted
relative to the radio synchrotron emission, where the bulk
relativistic Doppler factor $\delta$, is $(\Gamma(1-\beta
\cos{\theta}) )^{-1}$, with $\Gamma$ being the Lorentz factor of the
emitting region which is moving with a velocity $\beta$c at an angle
$\theta$ towards our line of sight. The spectral energy index is
$\alpha$, where flux density $\propto \nu^{-\alpha}$.
\citet{Tavecchio00} and \citet{Celotti01} showed how this effect could
explain the surprisingly large X-ray flux observed from the
PKS~0637-752 jet.

\clearpage

\begin{figure}[h]
%\begin{minipage}[c]{0.46\textwidth}
%  \includegraphics*[width=3.in]{f4.eps}
%{tavecBdelJ0841.ps}
\plotone{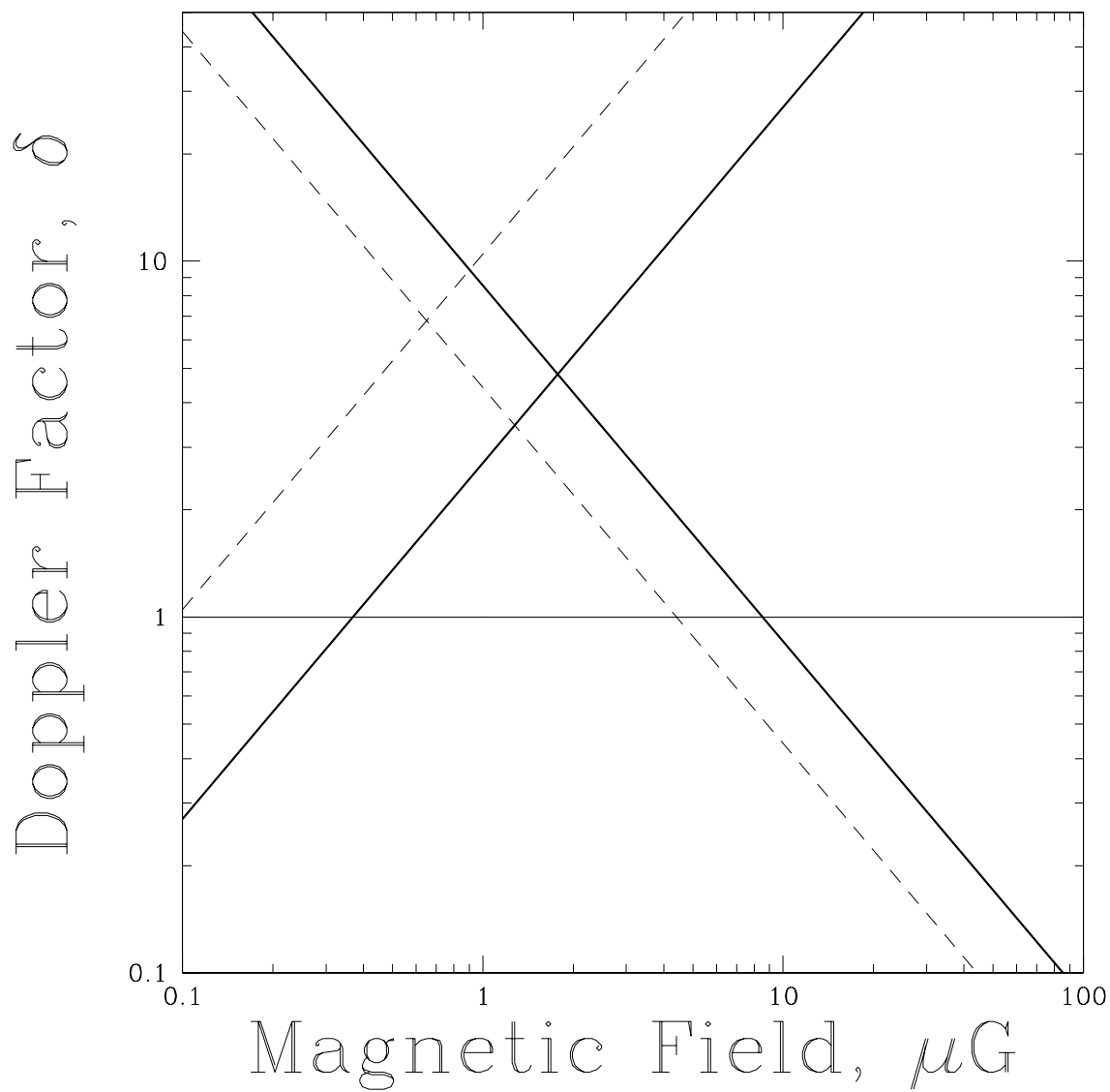}
\caption{\label{fig:tavec} Loci of equipartition ($\delta \propto$
  1/B) and of X-ray emission via IC/CMB ($\delta \propto$ B) in the
  cases that the 8.46 GHz flux of the jet is at its upper limit of 200
  $\mu$Jy (solid lines), or 10 times weaker (dashed
  lines). The intersection of solid (or dashed) lines gives a solution
  for the rest frame magnetic field and the Doppler factor.}
%\end{minipage}
\end{figure}

\clearpage

Figure~\ref{fig:tavec} applies the analysis of
\citet{Tavecchio00}. Here the lines with $\delta \propto$ 1/B show the
loci of equipartition between the magnetic fields and particles in the
jet rest frame.  We assume an electron population, n($\gamma$)$\propto
\gamma^{\rm {-m}}$, with spectral index m=2$\alpha$+1=2.4 producing
radio emission between 10$^6$ and 10$^{12}$ Hz, and with an equal energy
density in protons. We consider the emitting volume as a cylinder of
length 16\farcs3. We do not resolve the width of the cylinder, and
take the radius to be the 2\farcs1 semi-minor axis of the 62\%
encircled energy ellipse. The lines with $\delta \propto$ B show the
loci for which the same electron population giving the radio emission
produces the X-rays by IC/CMB. The intersection of the solid lines
give a solution for B and $\delta$ in the case that the jet flux is at
its limit of 200 $\mu$Jy at 8.46 GHz.  In that case, B = 1.7 $\mu$G
and $\delta$ =4.8. The magnetic field is an upper limit, and the
Doppler factor a lower limit, since the radio flux is just an upper
limit.  The lower limit to $\delta$ implies that the jet is within
12\arcdeg\ of our line of sight, and therefore at least 670 kpc in
length. For comparison, if f$_{\nu}$ were 20 $\mu$Jy, we would have
B=0.65 $\mu$G and $\delta$ = 6.8.  Since we do not resolve the jet, it
could be very much smaller. This would cause both B and $\delta$ to
have larger values than numbers quoted. In any case the  (B,$\delta$) point
must lie to the left and above the upward slanting solid line in
Figure~\ref{fig:tavec}, and to the right and above a line joining the
points where the two solid and two dashed lines intersect.

Electrons with $\gamma$ =  1000/$\Gamma$ produce $\approx$ 1 keV X-rays
when Compton scattering off the microwave background.  Such electrons will produce synchrotron radiation at too low a
frequency to be observed if B\,$\lesssim 10\, \Gamma^2 \mu$G. So an alternate
explanation for the observed lack of a radio jet is that the electron
spectrum breaks, e.g., due to ageing. If the radio break is at 1 GHz and
B=1.7$\mu$G, the electron spectrum breaks at a Lorentz factor $\le 10^4$.
The lifetime of $\gamma =10^4$ electons against Compton scattering on
the CMB at z=1.866 is about 3.6 $\Gamma^{-2}\, \times\, 10^6$ years.

\section{ALTERNATE INTERPRETATIONS}

Some faint galaxies, r$^{\prime}$=23 to 24, can be seen more or less
overlapping the region of the western X-ray contours in
Figure~\ref{fig:cont}. They are much too faint to expect that normal
galactic emission provides the X-rays, and the positions cannot be
associated with the X-ray emission peaks, especially after adjusting
the X-ray contours to coincide with the QSO.  Both these objections
could be overcome if these objects are a cluster of active galaxies.

Another possibility would be a foreground group of galaxies, at very
much lower redshift. This requires only a single unrelated source to
be superposed near the QSO by chance.  \citet{Bauer02} reports a
density of extended sources at this flux level to be $\approx$ 10
deg$^{-2}$, so there would be a 0.2\% chance of such a source at this
location.  Since the ChaMP project expects to study several thousand
sources, such a situation may occur.  However, it would be
strange that the X-rays do not center on the obvious z=0.32 galaxy
8\arcsec\ to the north.  The X-ray shape is quite distorted, so we
would be viewing  the cluster in an active and interesting dynamical
state. The cluster might be involved in gravitational lensing of the
QSO. We might have a failed cluster \citep{Tucker95} with only hot
gas and no galaxy formation.  In case of a foreground cluster, if hot gas overlaps
the QSO position future large throughput spectroscopy might use the
\citet{Krolik88} test to measure angular diameter distance
independently of redshift.  Any of these possibilities would result in 
J0841 being a very exciting system.

\section{CONCLUSIONS}

\citet{Schwartz02} has noted that X-ray emission by IC/CMB should result
in X-ray jets being cosmic beacons -- maintaining the same surface brightness
at any larger redshift.  This is because the (1+z)$^{-4}$ cosmic
diminution of surface brightness is exactly compensated by the 
(1+z)$^{4}$ increase in the energy density of the CMB with redshift. Such an
effect does not depend on equipartition, or on relativistic beaming.

The low magnetic field, $\le$2 $\mu$G, implied by the limits to
radio emission is unusual. Fields in clusters of galaxies can approach
1 $\mu$Gauss, while typical jet fields on kpc scales are of order 10
$\mu$Gauss. So the upper limits to magnetic field strengths derived
here are somewhat weak for a jet.  However, there seems to be no
fundamental physics prohibiting massive black holes to produce jets of
such low internal energy density.  Selection bias against finding
radio quiet X-ray jets could explain why such low magnetic field jets
have not previously been noticed. Alternately, this object may have a
magnetic field much weaker than the equipartition value.

\acknowledgments This work was supported in part by NASA contract
NAS8-39073 to the \emph{Chandra} X-ray Center, and CXC grants
AR2-3009X and GO2-3151C to SAO.  We thank D. Harris for discussions
and for comments on the manuscript, and D. Jerius for assistance with
telescope coordinate systems and the raytrace results. This research
used the NASA Astrophysics Data System Bibliographic Services, and the
NASA/IPAC Extragalactic Database (NED) which is operated by the Jet
Propulsion Laboratory, California Institute of Technology, under
contract with the National Aeronautics and Space Administration.  We
thank the VLA for the allocation of 1 hour of discretionary time. The
National Radio Astronomy Observatory is a facility of the National
Science Foundation operated under cooperative agreement by Associated
Universities, Inc.

\end{document}